\documentclass[conference]{IEEEtran}
\IEEEoverridecommandlockouts
\usepackage{cite}
\usepackage{amsmath,amssymb,amsfonts}
\usepackage{algorithmic}
\usepackage{graphicx}
\usepackage{textcomp}
\usepackage{xcolor}

\usepackage{times}
\usepackage{epsfig}
\usepackage{graphicx}
\usepackage{amsmath}
\usepackage{amssymb}
\usepackage{algorithmic}
\usepackage{textcomp}
\usepackage{subcaption}
\usepackage{balance}
\usepackage[ruled,vlined]{algorithm2e}
\usepackage[labelfont=bf]{caption}
\usepackage{multirow}
\usepackage{array}

\usepackage{lipsum}
\usepackage{multicol}

\def\BibTeX{{\rm B\kern-.05em{\sc i\kern-.025em b}\kern-.08em
    T\kern-.1667em\lower.7ex\hbox{E}\kern-.125emX}}

\begin{document}

\title{Blindness (Diabetic Retinopathy) Severity Scale
Detection}



\author{
\IEEEauthorblockN{Ramya Bygari, Rachita Naik}
\IEEEauthorblockA{\textit{Dept. of Computer Science \& Engg.} \\
\textit{National Institute of Technology Karnataka}, India \\
	ramyabygari239@gmail.com, rachitanaik590@gmail.com}
\and
\IEEEauthorblockN{Uday Kumar P}
\IEEEauthorblockA{\textit{Dept. of Information Technology} \\
\IEEEauthorblockA{\textit{National Institute of Technology Karnataka}, India
\\ uday.adroit1@gmail.com}
}
}

\maketitle

\begin{abstract}
Diabetic retinopathy (DR) is a severe complication of diabetes that can cause permanent blindness. Timely diagnosis and treatment of DR are critical to avoid total loss of vision. Manual diagnosis is time consuming and error-prone. In this paper, we propose a novel deep learning based method for automatic screening of retinal fundus images to detect and classify DR based on the severity. The method uses a dual-path configuration of deep neural networks to achieve the objective. In the first step, a modified UNet++ based retinal vessel segmentation is used to create a fundus image that emphasises elements like haemorrhages, cotton wool spots, and exudates that are vital to identify the DR stages. Subsequently, two convolutional neural networks (CNN) classifiers take the original image and the newly created fundus image respectively as inputs and identify the severity of DR on a scale of 0 to 4. These two scores are then passed through a shallow neural network classifier (ANN) to predict the final DR stage. The public datasets STARE, DRIVE, CHASE DB1, and APTOS are used for training and evaluation. Our method achieves an accuracy of 94.80\% and Quadratic Weighted Kappa (QWK) score of 0.9254, and outperform many state-of-the-art methods.
\end{abstract}

\begin{IEEEkeywords}
Diabetic retinopathy, fundus image, deep learning, convolutional neural network
\end{IEEEkeywords}

\section{Introduction}
\label{intro}
Diabetes causes hyperglycemia (high blood sugar). A broad range of biochemical changes is set-off within the body’s cells due to this condition that can result in structural and functional changes throughout the body, including the eye. Diabetic retinopathy (DR) is one such damage caused to the retina due to prolonged diabetic condition, which can eventually lead to vision loss. Globally, 629 million people in 20--79 age group are predicted to get affected with diabetes by the end of 2045, which is a 48\% increase from the current number~\cite{url01}. Diabetic Retinopathy is a serious non-communicable disease causing ocular morbidity, as per the Indian Health Ministry's National Diabetes and Diabetic Retinopathy Survey (2015--19)\cite{url02}.  
Studies reveal that over 90\% of visual loss from DR can be prevented with prompt curative intervention like eye tests and orderly screening at the initial stages~\cite{url07}. However, DR screening is handicapped by the lack of sufficient trained clinicians and intensive manual procedure. Moreover, the manual procedure is error-prone and time-consuming. The delay in procuring results or the inaccuracy in the diagnosis of DR can increase the risk of total vision loss.

\begin{figure}[t]
\begin{center}
   \includegraphics[width=.85\linewidth]{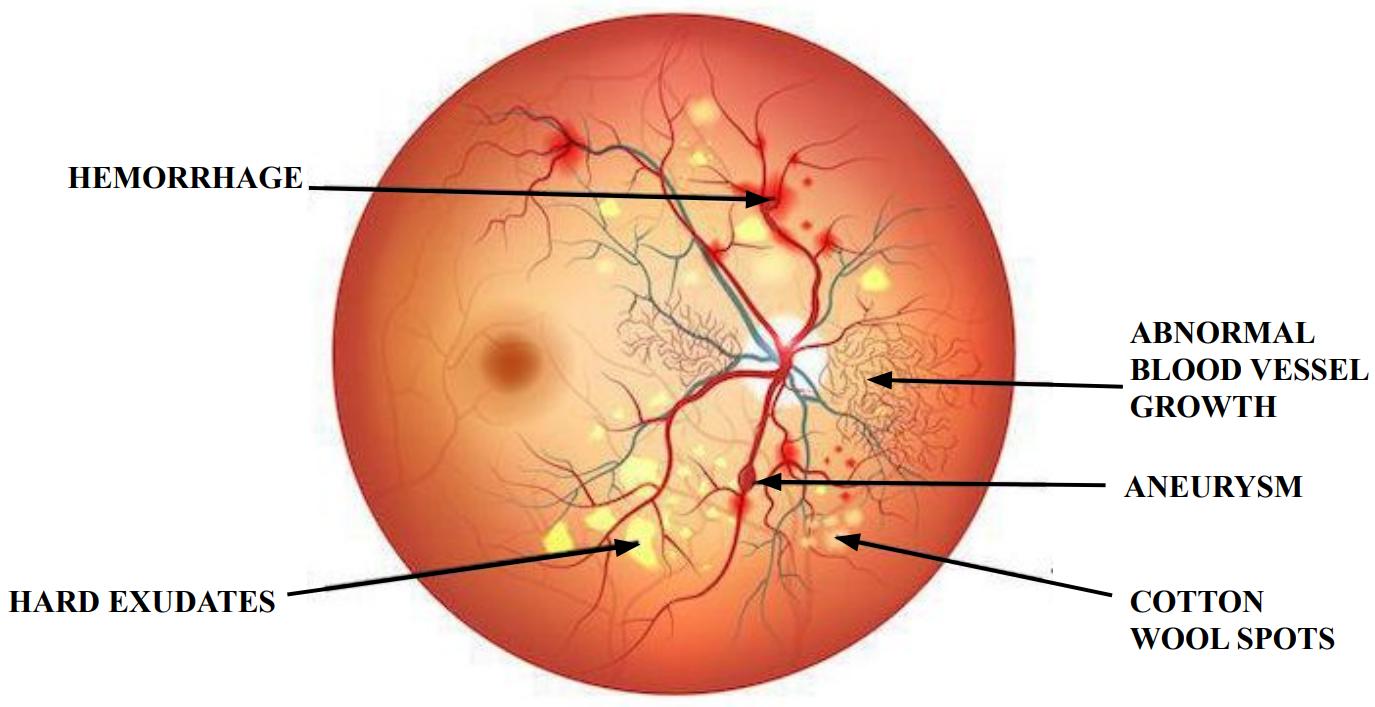}
\end{center}
   \caption{Microvascular complications of diabetic retinopathy (DR)}
   \label{fig1:DRComplications}
\end{figure}

\begin{figure*}[t]
\centering
\includegraphics[width=17cm,height=3cm]{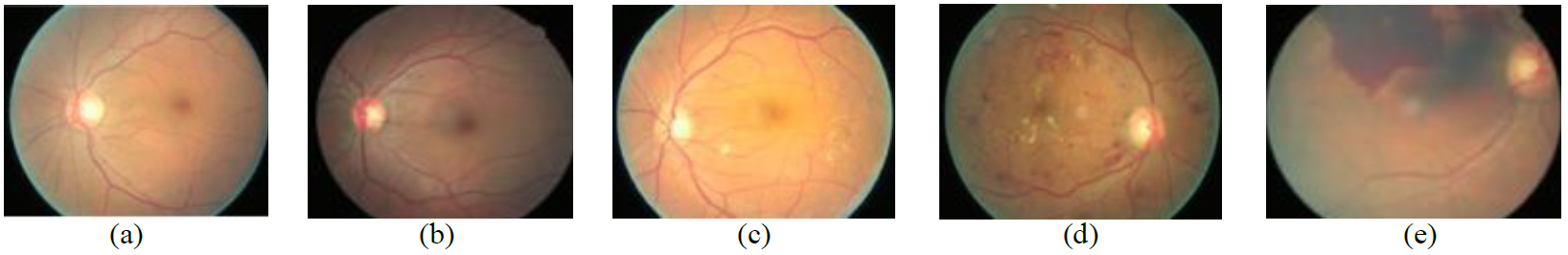}
\caption{Five severity stages of diabetic retinopathy (DR). (a) DR absent, (b) Mild level of DR, (c) Moderate DR, (d) Severe DR, (d) Proliferative level of DR}
\label{fig2:stageOfDR}
\end{figure*}
DR can hamper vision and ultimately lead to blindness due to the microvascular changes to the posterior section of the eye such as the development of microaneurysms, hard exudates, haemorrhages and macular oedema at the macula, and abnormal vessel growth (neovascularisation), that directly impact vision. The physical appearance of these complications is shown in Fig. \ref{fig1:DRComplications}.
The international standard for severity scale of DR~\cite{p02} has five stages such as: i) \textit{No detectable retinopathy}, ii) \textit{Mild \& non-proliferative retinopathy}, iii) \textit{Moderate \& non-proliferative DR}, iv) \textit{Severe non-proliferative retinopathy}, v) \textit{Proliferative retinopathy}. Retinal appearance in each of these stages is shown in Fig. \ref{fig2:stageOfDR}. Here, the disease progression is observed to be following the distinctive complications shown in Fig. \ref{fig1:DRComplications} that progressively increase in every stage. Hence, analysis of these features aids effective DR severity grade classification.

\par 
Initial solutions of computerized DR detection systems used hand-crafted features to train standard machine learning algorithms for grading DR. These features are sensitive to the focus angle, presence of artifacts, fundus image quality, and noise. Also, most of the approaches view the detection of DR as a two-class classification problem - DR or no DR. Binary classification does not provide sufficient information about the severity of the disease. These shortcomings necessitate the requirement of a deep learning system with automatic feature extraction for effective analysis of DR, and the classification of DR as a five-class problem (normal or No DR, mild level DR, moderate DR, severe DR, and proliferative level DR) rather than a two-class problem (DR or no DR). This will increase the efficiency and reproducibility of diagnostic tests. It will also allow doctors to make informed decisions about further testing and improve patient outcomes by providing appropriate treatment based on the disease stage.
\par
Convolutional neural networks (CNN) have produced superior performance in several recognition, classification, and analysis tasks in the medical imaging domain \cite{p18}. In this paper, we propose a novel CNN-based architecture for automatically predicting the severity stage of diabetic retinopathy, from human retinal fundus images. We modified UNet++ architecture for improved blood vessel segmentation from fundus images. Two EfficientNets are used for the prediction of the DR stage, one with the original fundus image and the other with the enhanced image. Finally, the two predictions are combined using an ANN to predict the final severity stage. The remaining sections of this paper are organised in the following way. Section \ref{lit-survey} presents a brief review of related works in literature, section \ref{proposed-method} explains the proposed methodology and the training approach in detail, and in section \ref{exp-results} the experimental results and related discussions are presented.




\section{Literature Survey}
\label{lit-survey}
A combined segmentation and classification pipeline has not been proposed before for DR detection. In this section, we discuss the existing methods in literature in the area of DR severity classification. The initial algorithms for the early classification of DR involved classic concepts from the fields of computer vision and machine learning. Acharya et al.~\cite{p03} extracted features like blood vessel area, microaneurysms, exudes, and hemorrhages which served as inputs to an SVM classifier to give an accuracy of 86\%, specificity of 86\%, and sensitivity of 82\% for five-class classification. Roychowdhury et al.~\cite{p04} presented a computer-aided diagnostic system DREAM for DR severity grading that consisted of a two-step hierarchical classification approach, achieving sensitivity -- 100\%, specificity -- 53.15\%, and AUC -- 0.903. However, the shortcoming in these types of approaches is the limited number of features used.\par
In the subsequent years, deep learning approaches demonstrated their supremacy over classic computer vision algorithms in object detection and classification tasks. Pratt et al.~\cite{p05} designed and developed a CNN architecture that could identify intricate features in the retina and consequently provide an automated diagnosis to classify the severity of DR, achieving a sensitivity of 95\% and an accuracy of 75\%. 
Lam et al.~\cite{p06} demonstrated the use of CNNs on color fundus images for DR staging, achieving a validation sensitivity of 95\%. They used contrast limited adaptive histogram equalization method for preprocessing and transfer learning with pre-trained GoogLeNet and AlexNet models and obtained test accuracies of 74.5\%, 68.8\%, and 57.2\% on 2-ary, 3-ary, and 4-ary classification models, respectively. 
Zhou et al.~\cite{p07} employed a multi-cell, multi-task CNN that combines cross entropy and mean square error (MSE) to classify images into 5 DR classes. Adly et al.~\cite{p08} proposed a binary tree based VggNet classifier that gave accuracy 83.20\%, sensitivity 81.80\% and specificity 89.30\%.~\cite{p09} used a pre-trained InceptionNet V3 for 5-class DR classification achieving an accuracy of 90.90\%. Sarki et al.~\cite{p10} used pre-trained ResNet50, XceptionNet, DenseNet, and VGG Net, achieving an impressive accuracy of 81.3\%. Both teams of researchers used datasets that were provided by APTOS and Kaggle ~\cite{url06}.

The above approaches do not take into consideration the effects of disease progression. Each stage of DR is characterised based on distinctive features like - microaneurysms, hard exudates, haemorrhages, and cotton wool spots that increase in every stage. These distinctive features can be captured by eliminating the retinal vessels, which is the novelty of our approach. 
Our automated system takes the fundus image as input and aims at segmenting the retinal vessels from the background with high accuracy to help detect abnormalities that aid the subsequent classification model to effectively categorise the DR. A novel retinal vessel segmentation model is presented in this work that addresses the multi-scale invariance and semantic gaps drawbacks. 

\section{Proposed method}
\label{proposed-method}
\begin{figure}[t]
\centering
\includegraphics[width=\linewidth]{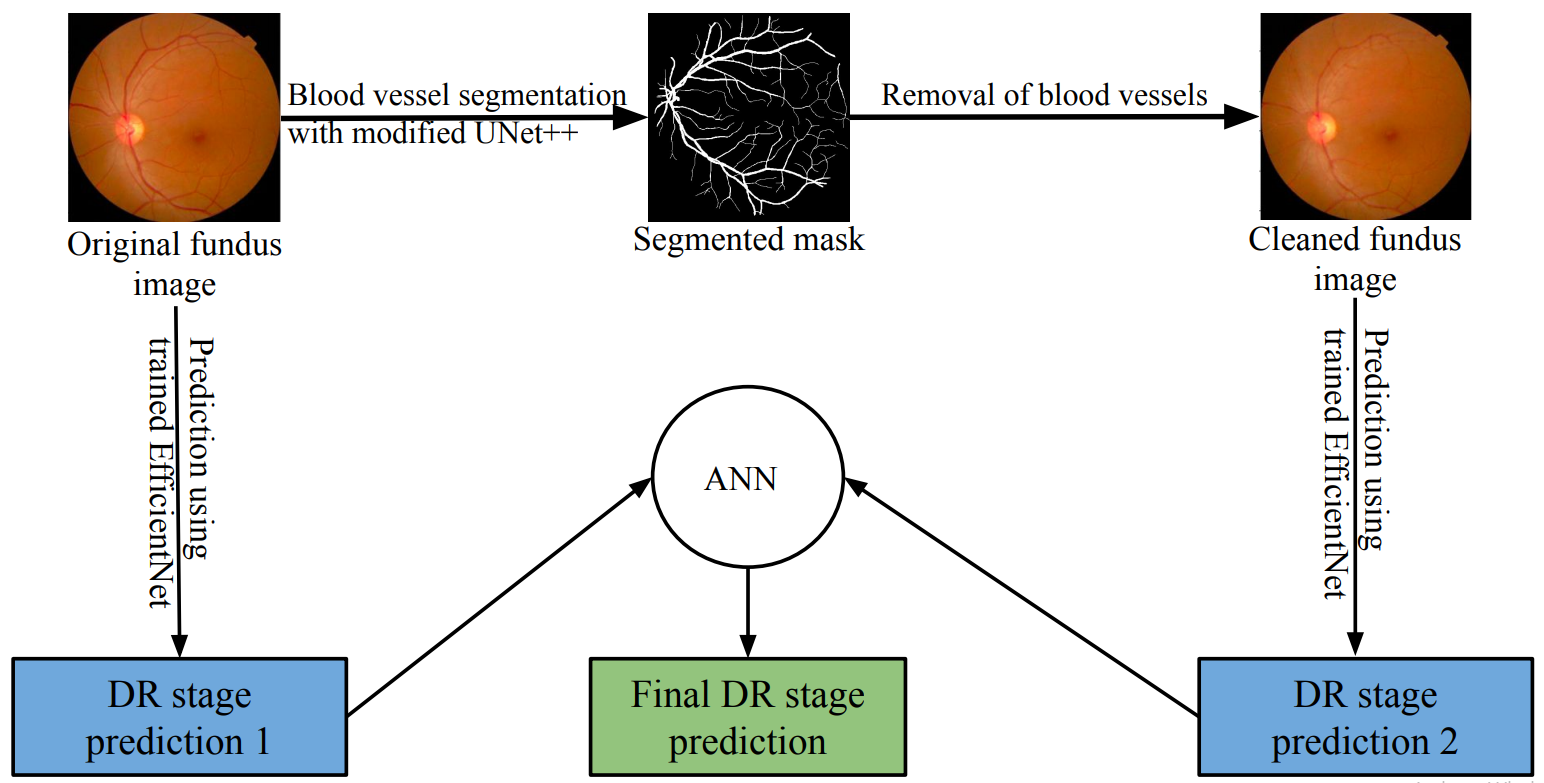}
\caption{Overview of the proposed DR severity detection method}
\label{fig:overview}
\end{figure}

The overview of the proposed method is shown in Fig. \ref{fig:overview}. It consists of three steps. The first step involves segmenting the retinal vessels from the fundus image. For this purpose, a novel retinal vessel segmentation model by modifying the basic UNet++ is used. This CNN model addresses drawbacks like the multi-scale invariance and semantic gaps. The segmented retinal vessel output (binary image) is then used to create an enhanced fundus image without retinal vessels while retaining cotton wool spots, exudates, and haemorrhages (if any), which are essential for retinopathy staging. The resulting image, i.e. the cleaned image, has distinct features for retinopathy staging emphasized. In the second step, the original and the cleaned images are fed into two separate EfficientNet~\cite{p11} classifier CNNs. Each EfficientNet outputs the grade corresponding to the severity of DR. Output from these two EfficientNets are passed to a simple ANN-classifier in the third step to obtain the final stage of diabetic retinopathy as output. Each of the three steps is further explained in detail. Since the entire pipeline comprises of three modules, each of them is explained separately in the following sections.

\subsection{Segmentation} 
\begin{figure}[t]
\centering
\includegraphics[width = 9cm,height= 5cm]{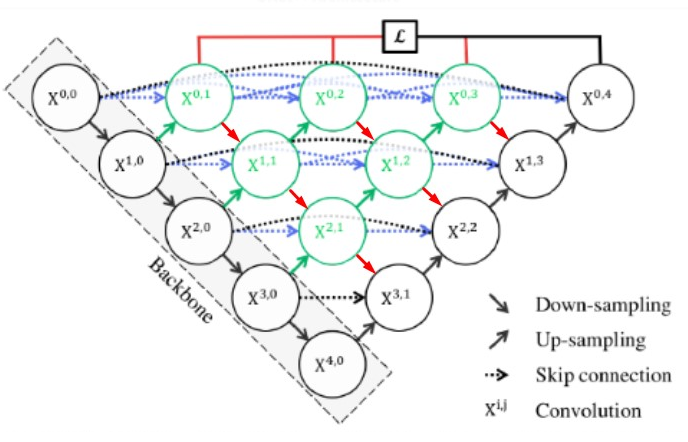}
\caption{Modified UNet++ architecture proposed}
\label{fig:modUnet}
\end{figure}

Base UNet~\cite{p12} architecture consists of an encoder (contraction path), to encapsulate the context in an image and a decoder, to capture accurate localization using deconvolution. Skip connections are introduced between the encoder and decoder to bridge the semantic gap between features extracted during expansion and contraction. UNet++~\cite{p13}, a superset of UNet, consists of redesigned skip pathways and dense connections inspired by DenseNet, that helps to preserve features at all semantic levels by allowing direct “communication” from one layer to another. It combines several UNets of varying depths with densely connected decoders at the same resolution to address two fundamental issues of UNet: the unknown depth of the optimal architecture and the unnecessarily restrictive design of skip connections. In this paper, a modified UNet++ architecture is proposed, that further improves segmentation accuracy while eliminating marginal segmentation errors. To allow the flow of information and bridge the semantic gaps between all levels in UNet++, the input to a node in the decoder layer consists of an additional downsampled output from the dense block above, fused with skip connections of the dense block in the same layer and the upsampled output from the lower dense block. This connects the feature maps between decoder nodes at different semantic levels and allows multiple paths for the flow of information. The architecture used for segmentation is shown in Fig. \ref{fig:modUnet}. The output of the segmentation is a binary image representing the retinal blood vessels.

\begin{algorithm}[b]
\KwIn{Channel, Mask}
\KwOut{Channel in cleaned image}
    \caption{{\bf To obtain Cleaned Image} \label{Algorithm}}
  \begin{algorithmic}
\STATE $blurImage\gets convolve(channel,filter)$ 
\STATE $blurImage\gets blurImage/filterSize$
\STATE $Channel[Mask==1]\gets blurImage$
  \end{algorithmic}
\end{algorithm}
The purpose of this module in the system is automated feature engineering. As mentioned in section \ref{intro}, microaneurysms, hard exudates, haemorrhages, and cotton wool spots progressively increase in every stage of DR. These features play an integral role in grading the severity of DR. Blood vessels are not required features for retinopa fthy staging. Hence we remove the blood vessels from the original fundus image with help of a binary vessel mask obtained from the segmentation step, to create an enhanced fundus image without blood vessels (cleaned image). This cleaned image devoid of blood vessels helps CNN in better learning of the characteristics of retinopathy-specific abnormalities.
\par
\begin{figure}[b]
    \includegraphics[width = \linewidth]{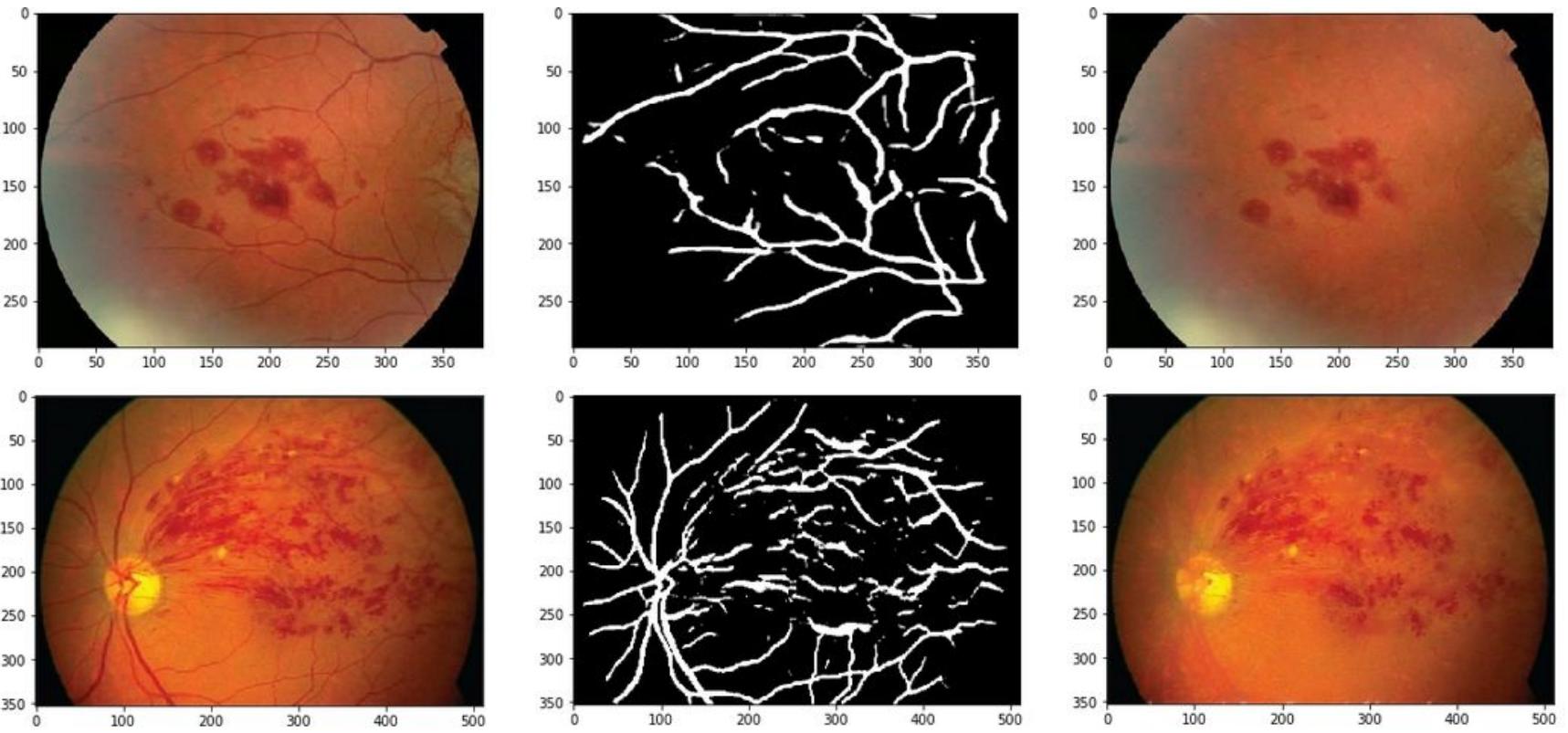}
    \caption{Cleaned fundus images (right) obtained using Algorithm \ref{Algorithm} that takes the original image (left) and segmented mask (center) as inputs}
    \label{fig:segresult}
\end{figure}

 The cleaned image is obtained using the Algorithm \ref{Algorithm}, which runs for every channel (R, G, B) of the fundus image. The inputs to the algorithm are the channels of the original fundus image and the segmented vessel mask. An intermediate blurred image is derived by convolving each original image channel with  4$\times$4, 16$\times$16, 32$\times$32, 64$\times$64 filters consecutively. If a pixel in the original image corresponds to a blood vessel pixel (i.e corresponding pixel value in the segmented mask is 1), then it is replaced with the pixel value in the intermediate image to give the final cleaned output.
The rightmost image shown in Fig. \ref{fig:segresult} is the cleaned fundus image free of the noisy blood vessel tree unlike the original fundus image (leftmost image). 



\par
The retinal images undergo the following prepossessing steps: The DRIVE~\cite{url03}, STARE~\cite{url04}, CHASE\_DB1~\cite{url05} datasets are combined to form one large dataset. Patches of size 256 $\times$ 256 are generated from each image. Approximately 6000 patches are generated. These patches are randomly shuffled to prevent any form of biased training before being fed into the network as input.

For segmentation, the modified UNet++ is used with Tversky loss~\cite{p01} as the loss function. This loss function combat the class imbalance problem (the number of background pixels \(>\!\!>\) number of blood vessel pixels). The modified UNet++ has a depth of 4. Stride 2 convolution is used for the encoder and stride 2 transposed convolution for the decoder. The number of filters (channels) starting with 32, doubles in every layer along the encoder. Adam optimizer (learning rate = 0.0001) is used for training. Dropout of 0.5, early Stopping, and L2 Regularisation have been used to prevent overfitting of the model. K-Fold cross validation  with K = 6 is used to evaluate the model. 1000 patches are set aside for testing in every fold.

\subsection{Intermediate stage predictions} 
\begin{figure}[t]
\centering
\includegraphics[width = \linewidth]{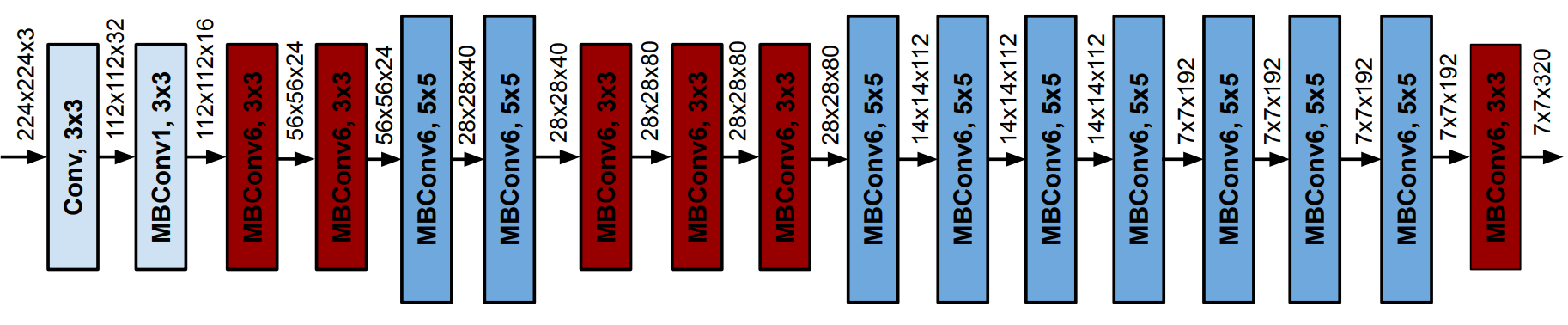}
\caption{Baseline EfficientNet architecture used in the proposed method}
\label{fig:effinet}
\end{figure}

Two EfficientNets (E1 and E2) are used in this stage. The input to E1 is the original fundus image to look at the characteristics of the fundus image as a whole. The input to E2 is the cleaned image without blood vessels produced in the previous stage. EfficientNet employs an efficient model scaling method to scale up each of the dimensions – depth, width, and resolution - of the CNN against the available resources in a more structured and principled manner. The basic architecture of the EfficientNet is shown in Fig. \ref{fig:effinet}. It uses mobile inverted bottleneck convolution (MBConv) as its building blocks, which resembles an inverted residual block used in MobileNetV2, combined with a depthwise separable convolution. Each subsequent model configuration in the family of EfficientNets refers to variants with more parameters and higher accuracy, that can be chosen depending on the resource availability.
\par
\begin{figure}[h]
    \includegraphics[width = 9cm]{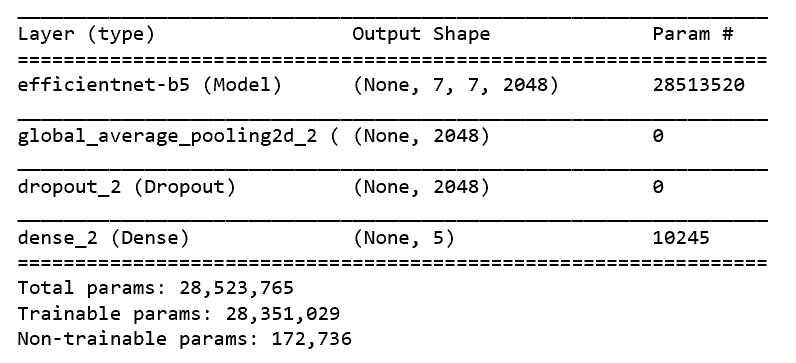}
    \caption{Summary of the pretrained model for DR Classification}
    \label{fig:modelsummary}
\end{figure}

 A pre-trained B5 version of EfficientNet has been used for modeling the two parallel paths of classification systems, with an input shape of ($224\times224\times3$). For the DR classification task, we require only five output classes. Hence, the final fully connected layer of EfficientNet (originally for 1000 classes) is customized to create a 5-class classifier. Subsequently, weights of the pre-trained EfficientNet-B5 model were placed into the transferred model. An average pooling layer with a 0.5 dropout is stacked on top of the EfficientNet CNN base model. The fully connected layer classifies the image into one of the five classes. Fig. \ref{fig:modelsummary} summarises the model used.
\par The results obtained with this parallel architecture, when compared with a single EfficientNet-based classification system, justify the choice for two separate classifiers. Having the E2 classifier score reduces the number of false negatives because of improved recognition of subtle features that characterise DR, thereby improving the sensitivity. We have retained E1, since in certain cases, it captures other subtle retinal vessel characteristics essential for DR grade detection that cannot be identified by E2.
\par

\subsection{Final stage prediction}
\begin{figure}[b]
\centering
    \includegraphics[width = 7cm,height=2.5cm]{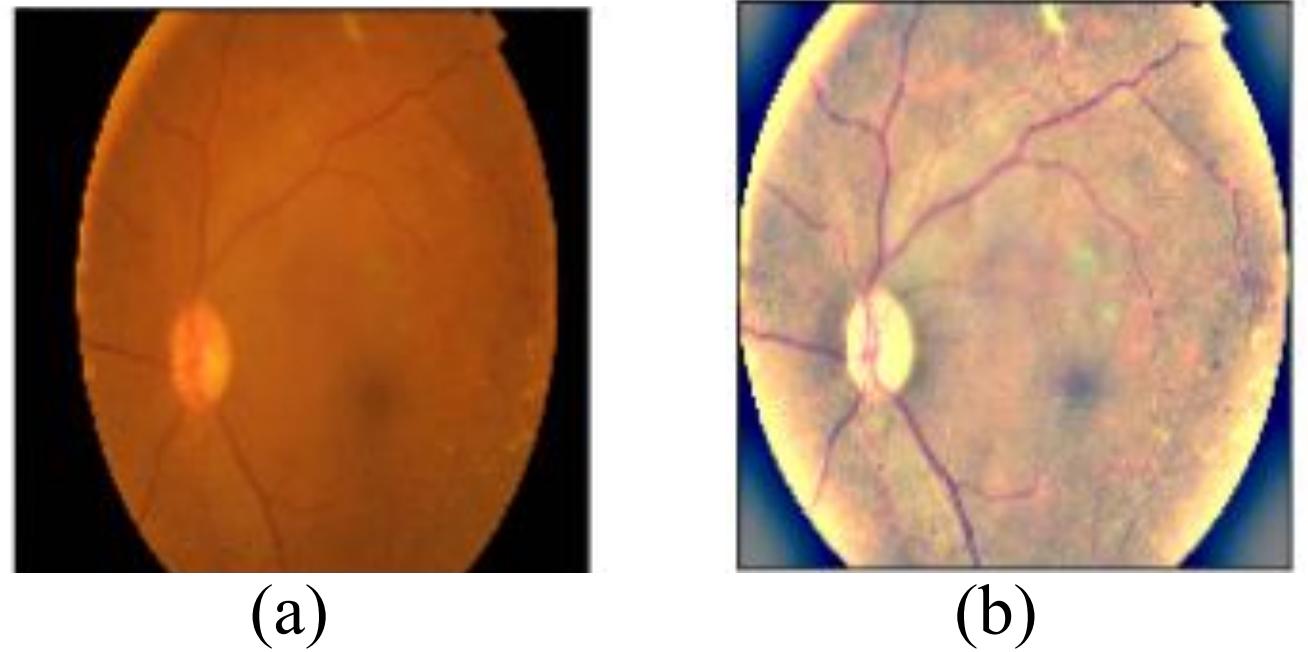}
    \caption{Preprocessing the fundus image, (a) Input image, (b) Preprocessed image}
    \label{fig:enhancedimg}
\end{figure}

A traditional ANN is used as the classifier for the final stage prediction. The inputs to this ANN are the output stage predictions of the two EfficentNets. The ANN consists of two hidden layers with three neurons each and uses ReLU as its activation function. The output layer uses the softmax activation function and has five neurons corresponding to the five stages of DR. The ANN is trained to map the output from the two EfficientNets to a final DR stage.
In the classification step, following preprocessing techniques are applied to the images of the APTOS dataset~\cite{url06} : 
\begin{itemize}
    \item To standardize the exposure and lighting variation, preprocessing steps suggested by Graham~\cite{r01}, are used. The images are re-sized to 224 $\times$ 224 to match the ImageNet dimensions, and the local average colour is subtracted and mapped to 50\% gray by adding Gaussian noise to increase robustness.
    \item An auto-cropping method is employed to remove the uninformative black regions. 
\end{itemize}
The final preprocessed image containing enhanced distinctive features is shown in Fig. \ref{fig:enhancedimg}(b).

Since our training images for the disease severity grading task are relatively small, on-the-fly augmentation is required to obtain a stable model with minimum overfitting. In order to obtain variance in the data, the major operations performed include -- rotation (between 0$^{\circ}$  to 360$^{\circ}$), shearing (angle between 20$^{\circ}$ and 200$^{\circ}$), flip (horizontal and vertical), zoom (random zoom with 0.15 range), and rescaling.

Instead of the conventional multiclass classification, using multilabel classification for this task slightly improved the metric scores. In the case of multilabel, if the target is a certain class, then it encompasses all the classes before it, e.g., class 4 retinopathy is normally encoded as [0, 0, 0, 0, 1], but in this case, it will be predicted as [1, 1, 1, 1, 1]. The method processes the output neurons in the order \textit{N\textsubscript{1}}, \textit{N\textsubscript{2}}, ...., \textit{N\textsubscript{k}}, and stops when a neurons output is below the predefined threshold (0.5), or there are no nodes left. The predicted category for the data point is the index \textit{k} of the last neurons \textit{N\textsubscript{k}}, whose output is larger than the threshold. The sigmoid activation function is applied to the model outputs, which are thresholded at 0.5 to find the position before the first zero.\par
Adam optimizer (learning rate = 0.00005) is selected for training along with the binary cross-entropy loss function. Early stopping with a patience of 12 epochs is used to counter overfitting. The learning rate is reduced if the loss on the validation set stagnates for four epochs. A small batch size of four is used to provide a regularizing effect and reduce generalization error. However, the model classification error is found to increase rapidly due to instability in batch normalization when the batch size is small. Hence, group normalization~\cite{p17} is used as a simple alternative in the EfficientNet layers. Group normalization divides the channels into groups, and then the mean and variance are computed for normalization within each group. This makes the computation independent of batch sizes, resulting in consistency of accuracy for different batch sizes. 

\section{Experiments and Results}
\label{exp-results}
\subsection{Dataset Description}
The DRIVE~\cite{url03}, STARE~\cite{url04} and CHASE\_DB1~\cite{url05} datasets have been concatenated and used for training the model for segmentation, since they contain fundus images captured under similar conditions. The DRIVE~\cite{url03} dataset is created using 400 diabetic patients aged between 25 and 90 years. 40 images have been randomly selected, out of which 33 do not show any sign of DR and 7 with signs of mild early DR. The CHASE\_DB1~\cite{url05} dataset contains 28 eye fundus images with a resolution of $1280 \times 960$. The STARE~\cite{url04} dataset consists of twenty images each of resolution $700 \times 605$, 10 of which are healthy and the other ten are diseased. 

The segmentation model, trained on the aforementioned data, is now used for obtaining predictions on the Kaggle APTOS 2019 challenge dataset~\cite{url06}, consisting of 3662 images, provided by the Aravind Eye Hospital, India. Each image has been rated for the severity of DR on an integer scale of zero(no DR) to four (proliferative DR). For the model testing, we selected 1000 images from the APTOS dataset~\cite{url06}. Both training and test sets contain an equal number of samples from all DR stages.

\subsection{Evaluation Metrics}
Quadratic Weighted Kappa (QWK)~\cite{p14}, also known as Cohen's Kappa, is used as the official scoring metric to evaluate the model for severity stage prediction of DR. The Kappa score is used to measure the agreement between two given ratings and is calculated between the scores assigned by a human evaluator and the values predicted by the model. It can range from 0 (the scores are completely different) to 1 (the scores are exactly alike). If two scores disagree, then the penalty will depend on how far apart they are. That means that Cohen's Kappa score will be higher if the ground truth is four, but the model predicts three, and the score will be lower if the model instead predicts a zero. Intuitively, a model that predicts severe retinopathy (three) when it is actually proliferative retinopathy (four) is better than a model that predicts mild retinopathy (one). \par
The rate of missed diagnoses is critical to clinical applications. Hence, the sensitivity and specificity of the models are also calculated. For these metrics, the two labels (no DR, mild DR) are considered negative, and the three DR labels (moderate, severe, and proliferative) are positive. When a sample belongs to the right class and is identified correctly, it is termed as either True Positive (TP) or True Negative (TN). False Positive (FP) refers to a negative sample incorrectly predicted as positive and False Negative (FN) means a positive sample incorrectly predicted as negative.

F1 score is the harmonic mean of precision and recall. F1 Score (Eq.~\ref{eq3}) is a popular metric for segmentation problems since it takes into account both false positives and false negatives, and it is a better indicator of performance when a class imbalance exists in the test set.

\begin{equation} 
\label{eq3}
\begin{split}
\text{F1 Score} = \frac{\text{2$\times$precision$\times$recall}}{\text{precision+recall}}
      = \frac{\text{2$\times$TP}}{\text{2$\times$TP+FP+FN}}\\
\end{split}
\end{equation}


\setlength{\tabcolsep}{6pt}
\begin{table}[tbh]
\label{tab:results}
\def\arraystretch{1.5}%
\centering
\begin{tabular}{|p{22mm}|p{12mm}|p{10mm}|p{10mm}|p{10mm}|}
\hline
\multicolumn{1}{|c|}{\textbf{Model}}
& {\textbf{F1 Score}}& {\textbf{Accuracy}} & \multicolumn{1}{|c|}{\textbf{AUC}}\\

\hline
Proposed method (modified UNet++) &  \textbf{0.8547} & 
\textbf{0.9754} & \textbf{0.9781} \\
\hline
Proposed method with UNet++ & 0.8205& 0.9568 & 0.9683 \\
\hline
IterNet\cite{Li_2020_WACV} & 0.8005 & 0.9478 & 0.9654 \\
\hline
DenseUNet\cite{p16} &  0.7996& 0.9423 & 0.9498\\
\hline
DUNet\cite{p15} &  0.8016 &0.9325 & 0.9391\\
\hline

\end{tabular}

\caption{Performance comparison of different architectures for fundus image segmentation with combined dataset}
\label{tab:segresult}
\end{table}

\setlength{\tabcolsep}{6pt}
\begin{table}[tbh]
\centering
\label{tab:results}
\def\arraystretch{1.5}%
\centering
\begin{tabular}{|p{22mm}|p{12mm}|p{10mm}|p{10mm}|}
\hline
{\textbf{\hspace{6mm}Dataset}}
& {\textbf{F1 Score}}& {\textbf{Accuracy}} & {\textbf{AUC}}\\

\hline
  CHASE\_DB1 & 0.8765 & 0.9832 & 0.9812 \\

\hline
  STARE & 0.8432 & 0.9581 & 0.9631 \\

\hline
 DRIVE & 0.8444 & 0.9849 & 0.9910 \\
\hline

\end{tabular}
\caption{Segmentation performance of the proposed method using DRIVE, STARE, CHARE\_DB1 datasets}

\label{tab:segresult}
\end{table}
\subsection{Performance Comparison}
\subsubsection{Segmentation}
The performance of our segmentation model is compared with IterNet~\cite{Li_2020_WACV}, DenseUNet~\cite{p16}, DUNet~\cite{p15}. K-Fold cross validation with K = 6 is performed on each of these models for comparison. It is observed that the proposed model outperforms state-of-the-art architectures. The final test results obtained on averaging over all folds are shown in Table \ref{tab:segresult}.

\subsubsection{Classification}
The performance of our two EfficientNets based classification system is compared against the pre-trained state-of-the-art CNN models - ResNet50, Inception-v3, and DenseNet along with the model proposed by Hagos et al.~\cite{p09} on Kaggle APTOS 2019 challenge dataset~\cite{url06}. Result of this comparison is presented in Table \ref{tab:classresults}. The model obtained an accuracy of 94.80\% with a sensitivity of 96.93\% and a specificity of 93.45\% on the test set. The system achieved a test QWK score of 0.92542.

\setlength{\tabcolsep}{6pt}
\begin{table}[t]
\def\arraystretch{1.5}%
\centering
\begin{tabular}{|p{32mm}|p{16mm}|p{16mm}|}
\hline
\multicolumn{1}{|c|}{\textbf {Model}}&\textbf {Accuracy}&\textbf {QWK Score}\\
\hline
Two EfficientNets (B5)& \textbf{0.9480} & \textbf{0.9254}\\
\hline
Single EfficientNet (B5)& 0.9390& 0.8965\\
\hline
ResNet50& 0.9217& 0.8886\\
\hline
Inceptionv3& 0.9373& 0.8700\\
\hline
DenseNet169& 0.9325& 0.8824\\
\hline
Hagos et al\cite{p09}.& 0.9091& 0.8996\\
\hline
\end{tabular}
\\
\caption{Peformance comparison for DR stage classification}
\label{tab:classresults}
\end{table}

\section{Conclusion}
In this paper, we presented a novel deep learning-based method for severity stage determination of diabetic retinopathy. We use a dual-path network of segmentation and classification CNNs in the process of classifying fundus images according to the severity level. The original fundus image is enhanced using a modified UNet++ CNN to emphasize the discriminative features required for retinopathy staging while eliminating unnecessary features. Further, both the images (original \& enhanced) are fed to two separate classifier CNNs (EfficientNets) to independently classify the DR. The final stage is determined by a simple ANN that takes the two independent predictions as the input. Our method pipeline shows the ability to learn the sophisticated features required to classify the fundus images based on severity. The results obtained indicate that the approach can provide reliable and consistent predictions without manual feature extraction and can serve as a computerized screening tool for early DR detection and staging. This fully autonomous model can assist ophthalmologists in making quick decisions. Experimental results show that our method performs better than many of the state-of-the-art methods.
\small
{
\balance
\bibliographystyle{IEEEtran}
\bibliography{egbib}

\begin{thebibliography}{10}
\providecommand{\url}[1]{#1}
\csname url@samestyle\endcsname
\providecommand{\newblock}{\relax}
\providecommand{\bibinfo}[2]{#2}
\providecommand{\BIBentrySTDinterwordspacing}{\spaceskip=0pt\relax}
\providecommand{\BIBentryALTinterwordstretchfactor}{4}
\providecommand{\BIBentryALTinterwordspacing}{\spaceskip=\fontdimen2\font plus
\BIBentryALTinterwordstretchfactor\fontdimen3\font minus
  \fontdimen4\font\relax}
\providecommand{\BIBforeignlanguage}[2]{{%
\expandafter\ifx\csname l@#1\endcsname\relax
\typeout{** WARNING: IEEEtran.bst: No hyphenation pattern has been}%
\typeout{** loaded for the language `#1'. Using the pattern for}%
\typeout{** the default language instead.}%
\else
\language=\csname l@#1\endcsname
\fi
#2}}
\providecommand{\BIBdecl}{\relax}
\BIBdecl

\bibitem{url01}
N.~Cho, J.~Shaw, S.~Karuranga, Y.~Huang, J.~da~Rocha~Fernandes, A.~Ohlrogge,
  and B.~Malanda, ``Idf diabetes atlas: Global estimates of diabetes prevalence
  for 2017 and projections for 2045,'' \emph{Diabetes research and clinical
  practice}, vol. 138, pp. 271--281, 2018.

\bibitem{url02}
S.~K. Mittal, P.~Nishant, A.~Agrawal, S.~Kumari, P.~Kumar, and A.~Chawhan,
  ``Community screening for diabetic retinopathy in uttarakhand, india, through
  targeted camps--a retrospective survey,'' \emph{Indian Journal of Community
  Ophthalmology}, vol.~1, pp. 19--21, 2020.

\bibitem{url07}
C.~J. Flaxel, R.~A. Adelman, S.~T. Bailey, A.~Fawzi, J.~I. Lim, G.~A.
  Vemulakonda, and G.-s. Ying, ``Diabetic retinopathy preferred practice
  pattern{\textregistered},'' \emph{Ophthalmology}, vol. 127, no.~1, pp.
  P66--P145, 2020.

\bibitem{p02}
C.~Wilkinson, F.~Ferris, R.~Klein, P.~Lee, C.~Agardh, M.~Davis, D.~Dills,
  A.~Kampik, R.~Pararajasegaram, and J.~Verdaguer, ``Proposed international
  clinical diabetic retinopathy and diabetic macular edema disease severity
  scales,'' \emph{Ophthalmology}, vol. 110, pp. 1677--82, 10 2003.

\bibitem{p18}
G.~Litjens, T.~Kooi, B.~E. Bejnordi, A.~A.~A. Setio, F.~Ciompi, M.~Ghafoorian,
  J.~A. Van Der~Laak, B.~Van~Ginneken, and C.~I. S{\'a}nchez, ``A survey on
  deep learning in medical image analysis,'' \emph{Medical image analysis},
  vol.~42, pp. 60--88, 2017.

\bibitem{p03}
A.~UR, L.~CM, N.~EY, C.~C, and T.~T, ``Computer-based detection of diabetes
  retinopathy stages using digital fundus images,'' \emph{Proceedings of the
  institution of mechanical engineers}, vol. 223, no.~5, pp. 545--553, 2008.

\bibitem{p04}
S.~{Roychowdhury}, D.~D. {Koozekanani}, and K.~K. {Parhi}, ``Dream: Diabetic
  retinopathy analysis using machine learning,'' \emph{IEEE Journal of
  Biomedical and Health Informatics}, vol.~18, no.~5, pp. 1717--1728, 2014.

\bibitem{p05}
H.~Pratt, F.~Coenen, D.~Broadbent, S.~Harding, and Y.~Zheng, ``Convolutional
  neural networks for diabetic retinopathy,'' \emph{Procedia Computer Science},
  vol.~90, pp. 200--205, 12 2016.

\bibitem{p06}
C.~Lam, D.~Yi, M.~Guo, and T.~Lindsey, ``Automated detection of diabetic
  retinopathy using deep learning,'' \emph{AMIA Joint Summits on Translational
  Science proceedings. AMIA Joint Summits on Translational Science}, vol. 2017,
  pp. 147--155, 05 2018.

\bibitem{p07}
K.~Zhou, Z.~Gu, W.~Liu, W.~Luo, J.~Cheng, S.~Gao, and J.~Liu, ``Multi-cell
  multi-task convolutional neural networks for diabetic retinopathy grading,''
  vol. 2018, 07 2018, pp. 2724--2727.

\bibitem{p08}
Adly, M.M., Ghoneim, A.S., Youssif, and A.A., ``On the grading of diabetic
  retinopathies using a binary-tree-based multiclass classifier of cnns.''
  \emph{International Journal of Computer Science and Information Security
  (IJCSIS)}, vol.~17, no.~1, 2019.

\bibitem{p09}
\BIBentryALTinterwordspacing
M.~T. Hagos and S.~Kant, ``Transfer learning based detection of diabetic
  retinopathy from small dataset,'' \emph{CoRR}, vol. abs/1905.07203, 2019.
  [Online]. Available: \url{http://arxiv.org/abs/1905.07203}
\BIBentrySTDinterwordspacing

\bibitem{p10}
R.~Sarki, S.~Michalska, K.~Ahmed, H.~Wang, and Y.~Zhang, ``Convolutional neural
  networks for mild diabetic retinopathy detection: An experimental study,''
  \emph{International Journal of Computer Science and Information Security
  (IJCSIS)}, 09 2019.

\bibitem{url06}
``Aptos kaggle dataset,''
  \url{https://www.kaggle.com/c/aptos2019-blindness-detection}.

\bibitem{p11}
M.~Tan and Q.~V. Le, ``Efficientnet: Rethinking model scaling for convolutional
  neural networks,'' \emph{CoRR}, vol. abs/1905.11946, 2019.

\bibitem{p12}
O.~Ronneberger, P.~Fischer, and T.~Brox, ``U-net: Convolutional networks for
  biomedical image segmentation,'' \emph{CoRR}, vol. abs/1505.04597, 2015.

\bibitem{p13}
Z.~Zhou, M.~M.~R. Siddiquee, N.~Tajbakhsh, and J.~Liang, ``Unet++: {A} nested
  u-net architecture for medical image segmentation,'' \emph{CoRR}.

\bibitem{url03}
``Drive dataset,'' \url{https://drive.grand-challenge.org/}, (Accessed on
  20/12/2020).

\bibitem{url04}
``Stare dataset,'' \url{https://cecas.clemson.edu/~ahoover/stare/}, (Accessed
  on 20/12/2020).

\bibitem{url05}
``Chase\_db1,'' \url{https://blogs.kingston.ac.uk/retinal/chasedb1/}, (Accessed
  on 20/12/2020).

\bibitem{p01}
S.~S.~M. Salehi, D.~Erdogmus, and A.~Gholipour, ``Tversky loss function for
  image segmentation using 3d fully convolutional deep networks,'' \emph{CoRR},
  vol. abs/1706.05721, 2017.

\bibitem{r01}
B.~Graham, ``Kaggle diabetic retinopathy detection competition report,'' Tech.
  Rep., 2015.

\bibitem{p17}
Y.~Wu and K.~He, ``Group normalization,'' \emph{CoRR}, vol. abs/1803.08494,
  2018.

\bibitem{p14}
J.~Cohen, ``Weighted kappa: Nominal scale agreement provision for scaled
  disagreement or partial credit,'' \emph{Psychological Bulletin}, vol.~70, pp.
  213--220, 4.

\bibitem{Li_2020_WACV}
L.~Li, M.~Verma, Y.~Nakashima, H.~Nagahara, and R.~Kawasaki, ``Iternet: Retinal
  image segmentation utilizing structural redundancy in vessel networks,'' in
  \emph{The IEEE Winter Conference on Applications of Computer Vision (WACV)},
  March 2020.

\bibitem{p16}
C.~Wang, Z.~Zhao, Q.~Ren, Y.~Xu, and Y.~Yi, ``Dense u-net based on patch-based
  learning for retinal vessel segmentation,'' \emph{Entropy}, vol.~21, p. 168,
  02 2019.

\bibitem{p15}
Q.~Jin, Z.~Meng, T.~Pham, Q.~Chen, L.~Wei, and R.~Su, ``Dunet: A deformable
  network for retinal vessel segmentation,'' \emph{Knowledge-Based Systems},
  vol. 178, 05 2019.

\end{thebibliography}
}

\end{document}